\documentstyle[preprint,epsf,prb,aps]{revtex}

\begin{document}

\title{ Mechanical Denaturation
of  DNA: existence of a low temperature denaturation}

\author{Enzo Orlandini$^1$,
Somendra M. Bhattacharjee$^2$ $^3$, 
Davide
Marenduzzo$^4$, Amos Maritan$^2$ $^4$  and 
 Flavio Seno$^1$}

\vskip 0.3cm
\address{ $1$ INFM - Dipartimento di Fisica ``G. Galilei'' , Via
Marzolo 8, 35100 Padova, Italy, \\
$2$ Abdus Salam International
Center for Theoretical Physics, Via Beirut 2, 34014 Trieste Italy,\\
$3$ Institute of Physics, Bhubaneswar 751 005 India,\\
$4$ International School for Advanced Studies,
 INFM, Via Beirut 2, 34014 Trieste Italy}

\maketitle


\begin{abstract}
  Recent theoretical predictions on DNA mechanical separation induced
  by pulling forces are numerically tested within a model in which
  self-avoidance for DNA strands is fully taken into account.  DNA
  strands are described by interacting pairs of self avoiding walks
  (SAW) which are pulled apart by a force applied at the two
  extremities.  The whole phase diagram is spanned by extensive Monte
  Carlo (MC) simulations and the existence of a low
  temperature denaturation is confirmed.
The basic features of
the phase diagram and the re-entrant phase boundary
are also obtained with a simple heuristic argument based on 
an energy-entropy estimate.   

\end{abstract}

\noindent 05.10.Ln, 64.60.Cn, 87.15.La, 87.14.Gy
\newpage

In the last few years, enormous technical improvements of experimental
devices such as atomic force microscopes \cite{Han96,Bus95}, optical
tweezers \cite{Svo94,Ash97} and glass microneedles\cite{Kis88,Clu96}
have allowed micromanipulation of single biological systems, and the
determination and characterization of new, in part unexpected,
behaviours of biomolecules\cite{Oes2000}.  These developments have
been particularly interesting and innovative for DNA molecules. DNA
micromanipulation techniques involve handles attached to the two ends
of the molecule which serve as anchoring points to manipulable
physical supports\cite{Strick98}.  In this way it has been possible to
study the response of DNA to external torques\cite{Strick99,Leger99}
and its mechanical unzipping in the absence of
enzymes\cite{Ess97,Bock97}. Nevertheless, only very recently
theoretical investigations have started to consider the denaturation of
DNA under the presence of forces or torques\cite{x1,x2,x3}.  On the
other hand theoretical studies of thermal denaturation (or melting) of
DNA have a much longer history\cite{PS66,x4}(for more recent studies see
\cite{x6,Causo})

Recently, Bhattacharjee\cite{Bat99} has extended a minimal
model\cite{PS66} in which the two strands of DNA are ideal polymer
chains, introducing a pulling force applied to two
extremities\cite{LN00,dyna}.  By mapping into a non-hermitian quantum
mechanics problem he has shown that the relative polymer can be pulled
out (i.e. the DNA can be unzipped) only if the applied force exceeds a
critical value.

These results have been further developed\cite{Maren1,Maren2} by
considering analytically solvable models in which DNA strands are
represented by directed walks. Surprisingly the force vs temperature
phase diagram revealed the presence\cite{Maren1,Maren2} of a novel
re-entrant unzipping transition at low temperature (see Fig. 1).
These results have been subsequently confirmed by other authors
\cite{Nelson,shak} within ideal chain models of DNA.

The aim of our paper is to study the phase diagram of DNA in the
presence of a pulling force with a more realistic model in which
self-avoidance for the DNA strands is explicitly taken into account to
verify the robustness of the low temperature re-entrance.

We model\cite{PS66,Causo} the two strands of a $2N$ base-pair DNA by
two interacting $N$ steps SAW, on a cubic lattice,
$\Gamma^{1}=\{\Gamma^{1}_0,\Gamma^{1}_1, \ldots, \Gamma^{1}_N\}$ and
$\Gamma^{2}=\{\Gamma^{2}_0,\Gamma^{2}_1, \ldots, \Gamma^{2}_N\}$, with
$\Gamma^{i}_p$ as the lattice site occupied by the $p$th monomer of
the $i$th chain ($i=1, 2$).  The first monomers $\Gamma^{1}_0,
\Gamma^{2}_0$ are kept fixed at one lattice unit apart. Each monomer
(site) on a strand corresponds to a base and it is supposed to have
its complementary at the same contour position in the other strand.
Two complementary monomers are favoured to be in contact (i. e., if
they occupy two nearest neighbour sites of the lattice) by an
energetic gain $\epsilon$ representing the binding energy.

In this approach\cite{PS66,Causo,Bat99,LN00} the sequence of bases is
not explicitly considered since the model is coarse grained in
character. In this respect each monomer is not to be thought of as a
single base, but as a group of bases (block). Consequently a mismatch
between corresponding blocks has to be very disfavoured with respect to
a correct matching. This aspect is adequately treated by our model.

To describe the mechanical unzipping of DNA an energetic term $-{\vec
  f} \cdot \vec{r}$ is added where $\vec{r}=\Gamma^1_N-\Gamma^2_N$ is
the vector joining the two ends of the strands and $\vec{f}$ is an
external force taken, without loss of generality, along the $z$ axis,
e.g. $\vec{f}= f \hat{e_z}$ (see Fig. 2).  Note that at $f=0$, this
model is in the same universality class as that of Ref. \cite{Causo}.

The Hamiltonian
of the problem  is then given by:
\begin{equation}
H_N = - \epsilon \left( \sum_{i=0}^N  
 \delta{\left(|\Gamma^1_i-\Gamma^2_i|,1 \right)} \right)
- \ \ {f}\hat{e}_z \cdot \vec{r},
\label{eq:eq1}
\end{equation}
where $\delta(a,b)$ is the Kronecker delta and $|x-y|$ is the
Euclidean distance between $x$ and $y$ in lattice units.  

\noindent Throughout all the
calculations $\epsilon = +1$ and therefore all the thermodynamic
properties of the systems depend on the temperature $T = \beta^{-1}$
and on the force $f$.  The free energy per monomer
${\cal{F}}_N(\beta,f)$, is

\begin{equation}
{\cal F}_N(T,f) = -\frac{1}{\beta N} \log{\left(
 \sum \exp{\left(-{\beta H_N}\right)  } \right)}
\label{eq:eq2}
\end{equation}
\noindent where the sum is over all the possible pairs of N-step $SAW$'s
$\Gamma^1$ and $\Gamma^2$.

The phase diagram can be studied by looking at the average fraction
number of contacts\cite{PS66} $ \Theta $ and at the z-component of the
average end point separation $\langle r^z \rangle$ defined as:
\begin{equation}
\Theta = \frac{\langle N_c \rangle}{N}
 =  - \frac{\partial}{\partial \epsilon } {\cal F}_N,
\ \ \
\langle r^z \rangle \sim \frac{\partial}{\partial f}\left( {\cal F}_N
N \right)
\label{eq:eq3}
\end{equation}

In the denaturated regime $\Theta$ should vanish whereas it should be
nonzero in the zipped phase.  To compute averages, for fixed values of
$f$ and $N$, Monte-Carlo simulations of the model have been performed
by a Metropolis based hybrid algorithm\cite{Madras,Orl} that consists
of local and global (pivot moves) deformations attempted on each
strand of the polymer. In addition a move that tries to interchange
the position of pairs of zipped (or unzipped) portions of the double
chain has been considered.

The efficiency of the algorithm has been further enhanced by the
multiple Markov chain  sampling scheme\cite{Tesi96,Orl}, 
a method that recently has been shown to be quite effective
in exploring the low temperature phase diagrams of interacting
polymers. All the runs have been performed by covering,
with at most $30$ different Markov chains, a large portion of the
temperature space ranging from $T=3$ down to $T=0.1$. The values of $N$
considered range from $2N=50$ up to $2N=150$ and the values of $f$ go
from $f=0$ up to $f=1$. 

In Fig.2 and Fig. 3 the numerical results respectively for $\Theta$
and $\langle r^z \rangle/ (2N)$ are plotted as a function of $T$ for different
values of the force strength $f$ and for systems of $2N=150$ monomers.
The data signal clearly the presence of three distinct regimes.

a) For $f < f_0 \sim  0.5$ the system undergoes a transition  from the
denaturated to the zipped regime. In particular at $f=0$
results similar
to those found in ref. \cite{Causo} are re-obtained, namely
a first order like transition at $T_D=  0.58 \pm 0.02$, where
however the fluctuation in the size of the ``bubbles'' diverges. 
Interestingly, our data support the prediction that the critical force
near the melting point ($T_D$) is $f \sim |T_D -T|^a$ with $a
<1$\cite{peliti,shak} (see also below).
This is different from  the 3-dimensional Gaussian chain  case
\cite{Bat99} or  the directed  
case of Fig. 1\cite{Maren1} (in $d=1+1$) where one finds $a=1$. In
these two cases the zero-force melting transition at $T_D$ is
continuous.

b) For $f_0 < f < f_t \sim 0.70$ the unzipped/zipped transition is
still present but if the temperature is further lowered the two
strands separate again through a low temperature denaturation or
unzipping induced by the presence of the force. This is established by
the peaks in Fig. 3 and by the minima in Fig. 4 for $f=0.625$ and
$f=0.65$.

Analysis of the specific heat peaks for different chain lengths
support the fact that both these transitions are first order.  One
might think of the existence of a subtle difference between the low
and the high temperature denaturation. To explore this possibility we
studied the size of the molten ``bubbles''.  For $f=0$ a simple
argument\cite{Causo} predicts that the size fluctuation of this
``bubble'' diverges when the thermal denaturation temperature $T_D$ is
approached from below.  As already mentioned we observed this effect
for $f=0$, but not for $f \neq 0$, both approaching the low
temperature transition from above or the high temperature transition
from below.  We explain this fact by noticing that for $f \neq 0$ the
transition temperature is less than $T_D$ and the part of the two
strands containing the ``bubbles'' is not influenced by the presence
of the external force (the external force act as a boundary effect).
Thus, by approaching the transition line from the zipped phase,
bubbles can never become critical if $f\neq{0}$. Therefore we conclude
that from a statistical point of view low and high temperature
denaturation transition are equivalent.

c) For $f > f_t$ the pulling force is always dominating over the
base-base interaction and the system is always in the denaturated
state.
 
The value of $f_0$ can be exactly calculated and the basic features of
the phase diagram and the re-entrant phase boundary
can be obtained with a simple heuristic argument based on 
an energy-entropy estimate.   Indeed at $T=0$, when
thermal fluctuations are absent, the only allowed configurations are
those in which the first $l$ monomers of the strands are parallel and
form $l$ contacts, whereas the remaining $N-l$ are pointing in
opposite directions parallel to the force.  Hence, the resulting
energy is a simple function of $l$:
\begin{equation}
E(l)=-l(1-2f)-f(2N+1)
\label{eq:eq4}
\end{equation}
For $f > 1/2$ the energy minimum occurs for $l=0$ (complete
stretching) whereas for $f < 1/2$ it occurs for $l=N$ (complete
zipping).  This argument states definitively that $f_0=1/2$ in full
agreement with our numerical findings. Notice that for $f=f_0=1/2, \ 
E(l)$ is independent of $l$, indicating coexistence.

For low enough temperatures we expect that the dominant
configurations are still of   $Y$-like configurations with $2(N-l)$
monomers zipped and two ``free'' strands each of length $l$ 
whose extrema are kept at a relative position $\vec{r}$ 
by the force $\vec{f}$.  
In this Y model the energy is thus given by:
\begin{equation}
E=-(N-l)\epsilon -{\vec r}\cdot {\vec f}
\label{eq:amos1}
\end{equation}
whereas the entropy is
\begin{equation}
S=\log{ \left[ P_{2l}(\vec{r}) \mu_{u}^{2l} \mu_z^{N-l} \right]}
\label{eq:amos2}
\end{equation}
where $P_N(\vec{r})$ is the probability that an $N$-step strand 
has an end-to-end displacement $\vec{r}$ and $\mu_{u}$ and $\mu_z$
are the effective coordination of the single strand and of the double
strand respectively. For the simple model introduced before 
 $\mu_z=\mu_u$ but in general this need not be true.

It is well known\cite{MEF} that
\begin{equation}
P_{2l}(\vec{r}) \sim \exp \left[-\left(\frac{r}{l^{\nu}}\right)^{\delta} 
c \right ]
\ \ \ \ \ l^{\nu} << r << l
\label{amos3}
\end{equation}
\noindent where $r=|\vec r|$, $\delta = (1 - \nu)^{-1}$, $\nu$
is the usual polymer critical exponent\cite{MEF}
 and $c$ is a constant. For
an ideal polymer $\nu = 1/2$ and one recovers the usual Gaussian
distribution.

In the last equation power law corrections lead to subleading terms in
the entropy estimate. From the last three equations the following free
energy is obtained:
\begin{equation}
F(l,\vec{r}) = l ( \epsilon - T \log{\frac{\mu_u^2}{\mu_z}}) - \vec{r}
\cdot \vec{f} + c T \left(  \frac{r}{l^{\nu}} \right)^{\delta} + {\rm const}
\label{amos4}
\end{equation}

Minimization of $F$ with respect to $\vec{r}$ yields  
\begin{equation}
r(l,f)= l \left( \frac{f}{c \delta T}\right)^{\frac{1}{\delta-1}},  
\label{amos5}
\end{equation}
where $r$ is the magnitude of $\vec{r}$ in the direction of $\vec f$. This $r$ gives a 
free energy
\begin{equation}
F(l) \equiv F (l,r(l,f)) = l \left[ \epsilon -
  T\log{\frac{\mu_u^2}{\mu_z}} 
- \frac{\delta -1}{ \delta^{\frac{\delta}{\delta-1}}}(cT)^{-\frac{1}{\delta-1}}
f^{\frac{1}{\nu}}  \right]
\label{eq:amos6}
\end{equation}
implying that $F(l)$ gets its minimum at $l=0$ (zipped phase) and
$l=N$ (unzipped phase) when $f<f_c(T)$ and $f > f_c(T)$ respectively
with
\begin{equation}
f_c(T) \sim T^{1 - \nu} \left( 1 - \frac{T}{T_D} \right)^{{\nu}},
\label{amos7}
\end{equation}
$T_D= {\epsilon}/{  \log\left({\frac{\mu^2_{u}}{\mu_z}
}\right)}$ being the thermal
denaturation temperature for this Y model.
Thus at $f_c(T)$ a first order phase transition  occurs where the
fraction of the bounded bases has a jump.
The last equation predicts a re-entrance with $f_c \to 0$ when $T \to
0$. This limit is given by a spurious effect due to the assumption
that Eq. \ref{amos3} applies also at very low $T$ where eq. \ref{amos5}
predicts $r \to \infty$. Indeed if Eq. \ref{amos3} is assumed to hold
till $P_{2l}(r) \mu_u^{2l} \sim 1$, e.g. $ r\leq r_M \sim 2 l$, and
after that $r$ remains constant, then   we easily find
that at low T 
\begin{equation}
f_c(T) = \frac{1}{2} ( \epsilon + T \log{\mu_z}).
\label{amos8}
\end{equation}
Eq. (\ref{amos7}) near the denaturation temperature and eq.
(\ref{amos8}) at the lowest $T$ predict a behaviour qualitatively in
agreement with the one described in fig. 1\cite{somen} and confirmed by our
simulation analysis.

In conclusion we have shown how a pulling force applied to the extrema
of a model of double stranded DNA molecule could introduce non trivial effects in the phase
diagram.
The results  obtained on MC simulations  in which, unlike previous
studies, excluded volume effects  are fully taken
into
account, confirm the existence of a low temperature denaturation.
 Although in our approach the intrinsic helicity of the DNA
and the randomness of the bases sequences have not been incorporated,
we believe , on the basis of preliminary numerical results,
that they should not affect the main features of the  phase diagram and
in particular the existence of a re-entrance region. Therefore we
expect that such an effect
could be pinpointed by high precision measurements of DNA denaturation
under a pulling force.

\medskip
{\bf Acknowledgments}: We thank A. Trovato
for stimulating discussions. FS acknowledges hospitality by
SISSA and ICTP.  This work was supported by MURST (COFIN 99).



\begin{figure}
\begin{center}
\epsfxsize=6.4in
\epsfbox{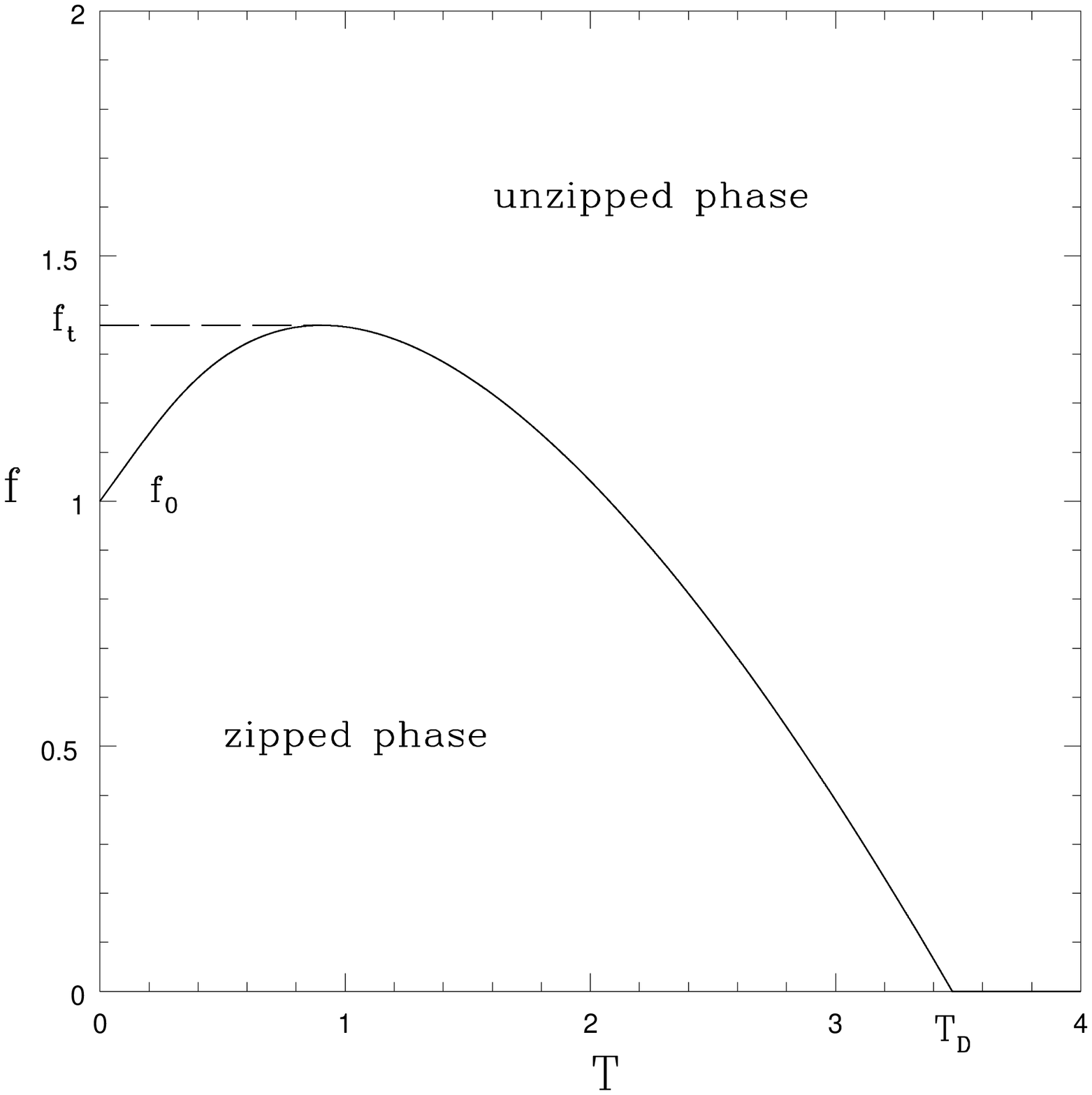}
\end{center}
\caption{The exact force (f) versus
 temperature (T) phase diagram as 
obtained in ref. \cite{Maren1} for a directed model (d=2) of DNA.
See eq. (1) for definition of force. The values of $f_t$, $f_0$ and
$T_D$ are model dependent.} 
\label{fig:histo}
\end{figure}

\begin{figure}
\begin{center}
\epsfxsize=4.4in
\epsfbox{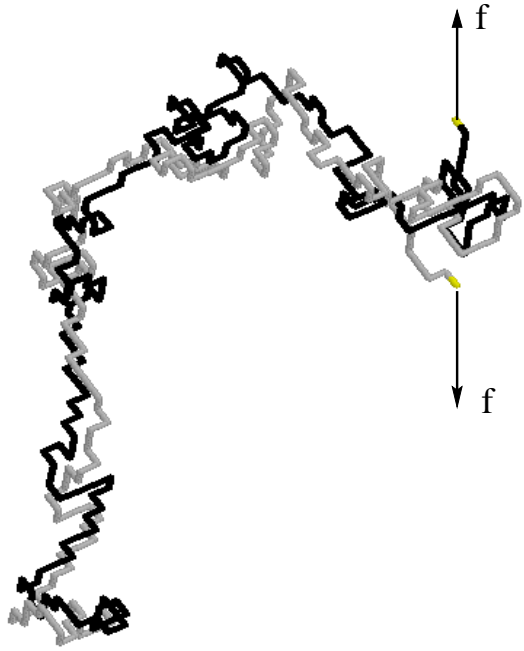} 
\end{center}
\caption{A typical configuration of a double stranded DNA
chains $(2N=200)$ obtained with the model described in the text.
The different colours indicate the two different strands.}
\label{fig:fig1}
\end{figure}

\begin{figure}
\begin{center}
\epsfxsize=6.4in
\epsfbox{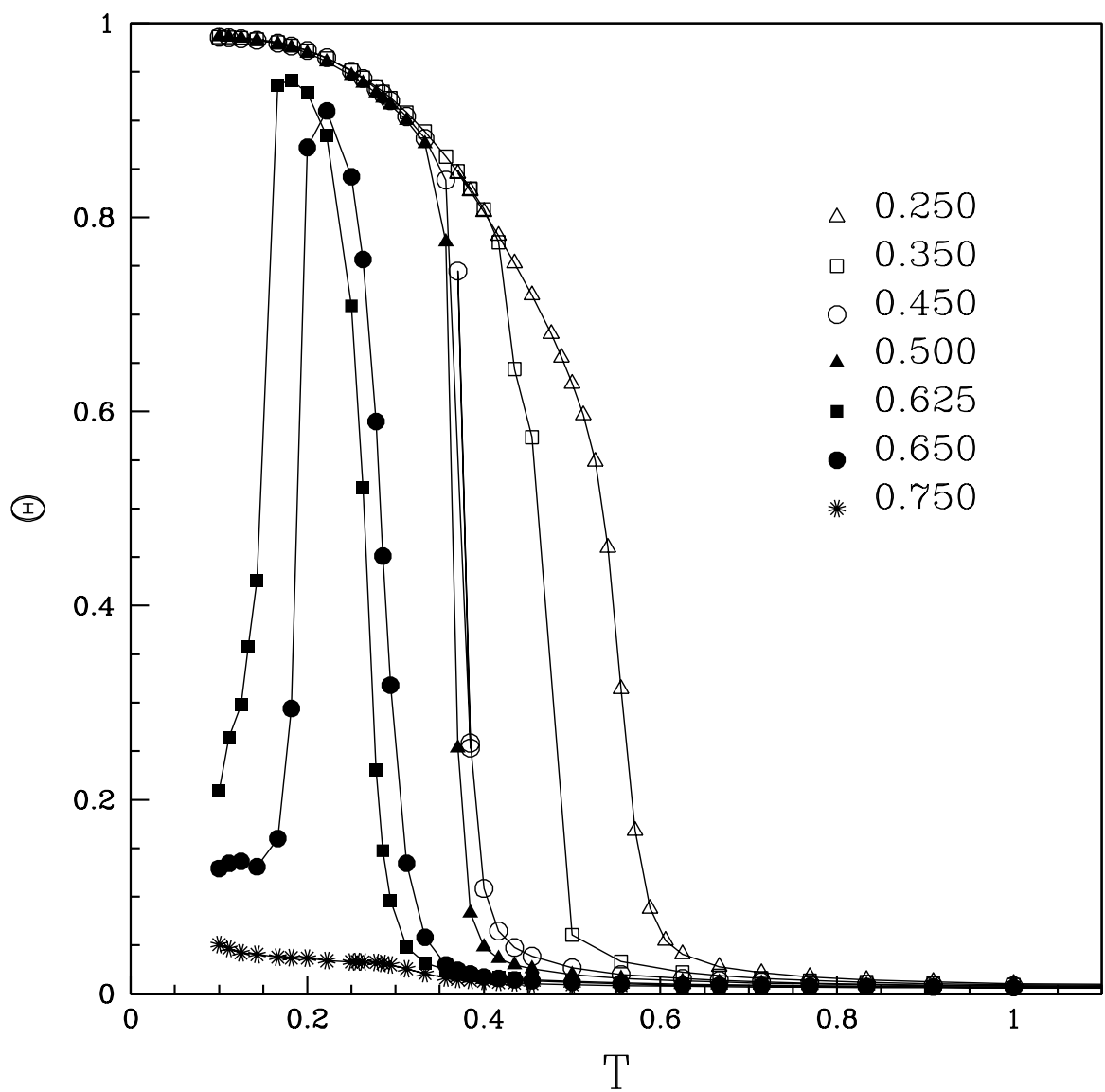}
\end{center}
\caption{Average fraction of bound pairs $\Theta$ versus T, for several
values of the force $f$ and for $2N=150$. Error bars are of the size
of the symbols. The two peaks ($f=0.625$ and
$f=0.65$) signal a zipped phase between two unzipped regions (re-entrance).}
\label{fig:fig2}
\end{figure}

\begin{figure}
\begin{center}
\epsfxsize=6.4in
\epsfbox{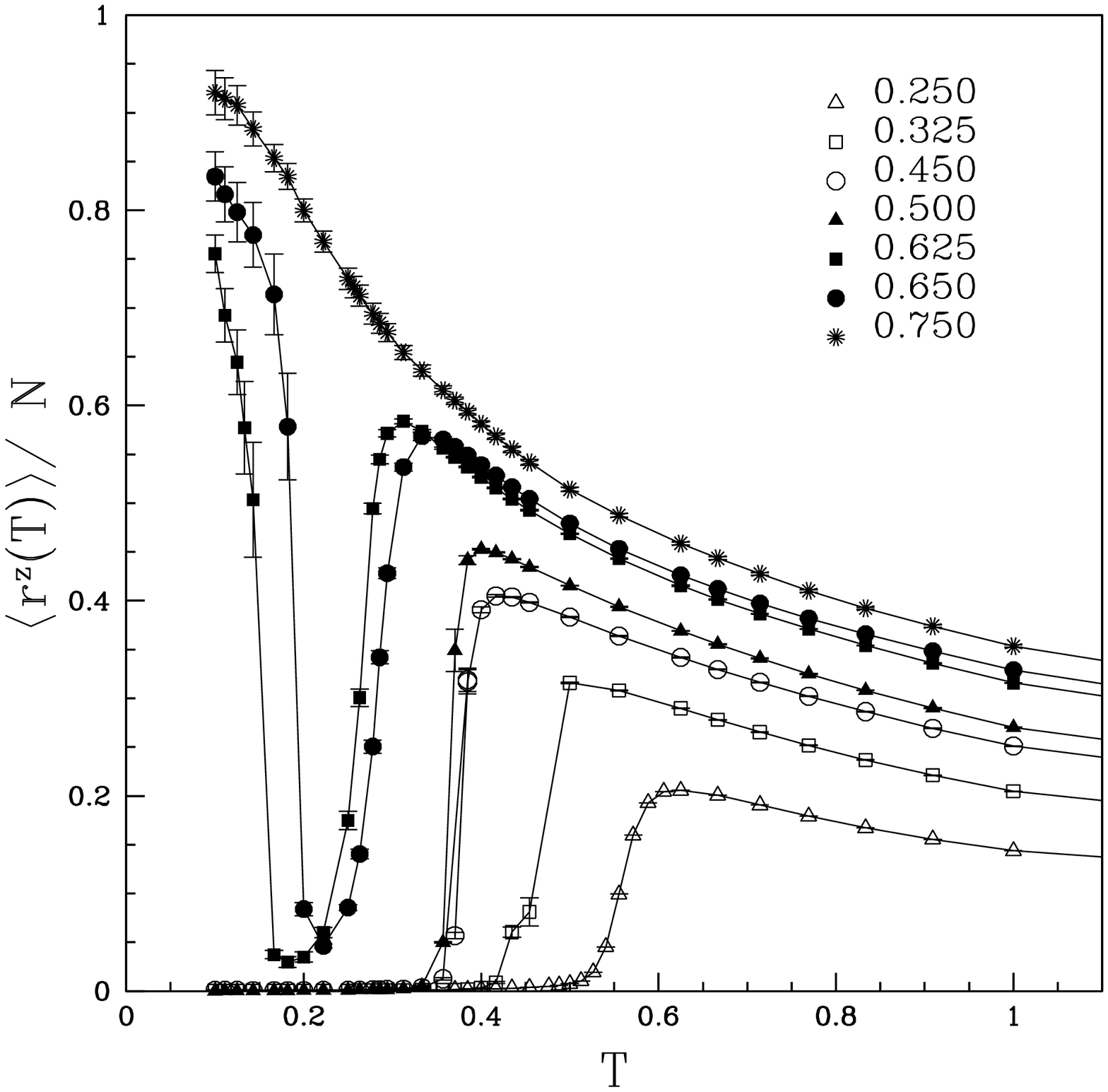}
\end{center}
\caption{End to end end separation in the force direction versus T,
for 
different values of the force $f$ for $2N=150$. Error bars correspond
to one standard deviation. 
The two minima ($f=0.625$ and
$f=0.65$) signal a zipped phase between two unzipped regions (re-entrance).
} 
\label{fig:fig3}
\end{figure}


\begin{thebibliography}{30}

\bibitem{Han96} H.G. Hansma, J. Vac. Sci. Technol. {\bf B14}, 1390 (1996).

           
\bibitem{Bus95} C. Bustamante \& D. Keller, Phys. Today {\bf 48}(12),
  32 (1995).

\bibitem{Svo94} K. Svoboda \& S.M. Block,
  Annu. Rev. Biophysics. Biomol.
Struct. {\bf 23}, 247 (1994).



\bibitem{Ash97} A. Ashkin, Proc. Natl. Acad. Sci. USA {\bf 94}, 4853
  (1997).

\bibitem{Kis88} A. Kishino \& T. Yanagida, Nature {\bf 34}, 74 (1988).


\bibitem{Clu96} P. Cluzel, A. Lebrun, C. Heller, R. Lavery,
  J.L. Viovy,
D. Chatenay \& F. Caron, Science {\bf 271}, 792 (1996).

\bibitem{Oes2000} 
 V. Moy , E.L. Florin  \& H.E. Gaub , Science {\bf
264}, 257 (1994); 
R. Merkel, A. Leung, K. Ritchie \&  E. Evan, Nature {\bf 397}, 50 (1999);
M. Rief, F. Oesterhelt, B. Heymann \& H.E. Gaub, 
Science {\bf 276},1109 (1997).

\bibitem{Strick98} T.R. Strick, J.F. Allemand, V. Croquette \&
  D. Bensimon,
Journal of Statistical Physics {\bf 93}, 647 (1998).


\bibitem{Strick99} T.R. Strick, J.F. Allemand, V. Croquette \&
  D. Bensimon,
Physica {\bf 263A}, 392 (1999).

\bibitem{Leger99} J.F. Leger, G. Romano, A. Sarkar, J. Robert,
  I. bordieu, D Chatenay \& J.F. Marko, Phys. Rev. Lett. {\bf 83},
  1066 (1999).

\bibitem{Ess97}   B. Essevaz-Roulet, U. Bockelmann \& F. Heslot, Proc.
Natl. Acad. Sci. USA {\bf 94}, 11935 (1997).

\bibitem{Bock97} U. Bockelmann, B. Essevaz-Roulet \& F. Heslot, Phys.
Rev. Lett. {\bf 79}, 4489 (1997); Phys. Rev. {\bf E58}, 2386 (1998).


\bibitem{x1}  S. Cocco \& R. Monasson, Phys. Rev. Lett. {\bf 83} 5178
(1999).

\bibitem{x2} M. Peyrard, Europhys. Lett. {\bf 44}, 271 (1998).

\bibitem{x3} R.E. Thompson \& E. Siggia, Europhys. Lett. {\bf 31},
335 (1995).


\bibitem{PS66} D. Poland \& H.A. Scheraga. 
 J. of Chem. Phys. {\bf 45}, 1464-1469 (1966).


\bibitem{x4} M.E. Fisher, J. Chem. Phys. {\bf 45}, 1469 (1966).

\bibitem{x6} D. Cule \& T. Hwa, Phys. Rev. Lett. {\bf 79},
2375 (1997).

\bibitem{Causo} M.S. Causo, B. Coluzzi \& P. Grassberger,
 Phys. Rev. {\bf E62}, 3958  (2000).

\bibitem{Bat99} S.M. Bhattacharjee, cond-mat/9912297, J. Phys.
{\bf A 33} L423 (2000); J. Phys. {\bf A 33}, 9003 (2000);
cond-mat/0010132.

\bibitem{LN00} More recently a similar model which takes into account
also the presence
of sequence disorder has been studied by D. K. Lubensky \&
D. R. Nelson, Phys. Rev. Lett. {\bf 85}, 1572
  (2000).  See also S. M. Bhattacharjee and D. Marenduzzo,
  cond-mat/0106110.



\bibitem{dyna} The dynamics of unzipping has also been studied 
recently: see \cite{Maren1}, 
K. L. Sebastian, Phys. Rev. E {\bf 62},
   1128 (2000); S. Cocco, R. Monasson and J. F. Marko,
   Proc. Natl. Acad. Sci. USA {\bf 98}, 8908 (2001).


\bibitem{Maren1} D. Marenduzzo, S.M. Bhattacharjee, A. Maritan,
E. Orlandini and F. Seno, cond-mat/0103142.


\bibitem{Maren2} D. Marenduzzo, A. Trovato and A. Maritan,
Phys. Rev. E {\bf 64}, 031901 (2001).

\bibitem{Nelson} D. K. Lubensky, D. R. Nelson, cond-mat/0107423.

\bibitem{shak} E. A. Mukamel, E. I. Shakhnovich, cond-mat/0108447.







\bibitem{Madras} N. Madras \& A. Sokal, J. Stat. Phys. {\bf 47}, 573 (1987).

\bibitem{Orl} E. Orlandini, F. Seno \& A.L. Stella,
  Phys. Rev. Lett.
{\bf 84}, 294 (2000).

\bibitem{Tesi96} M.C. Tesi, E.J. van Rensburg, 
E. Orlandini \& S.G. Whittington, J. Stat. Phys. {\bf 29}, 2451
(1996).

\bibitem{peliti} Y. Kafri, D. Mukamel, L. Peliti, cond-mat/0108323.


\bibitem{MEF} M. E. Fisher, J. Chem. Phys. {\bf 44}, 616 (1966), see also
P.G. de Gennes, {\em Scaling concepts in polymer physics} (Cornell Univ. Press, Ithaca, 
1979) sec. 1.2.

\bibitem{somen} In the critical case of fig. 1 $\nu$ is the critical
correlation length exponent for melting as in Ref.\cite{Bat99}


\end{thebibliography}
\end{document}